# Using ChatGPT for Science Learning : A Study on Pre-service Teachers' Lesson Planning

Gyeong-Geon Lee, and Xiaoming Zhai (AI4STEM Education Center, University of Georgia, Athens, GA 30602, United States)

*Abstract*— While there have been efforts to integrate cutting-edge technologies into science teaching and learning, ChatGPT has emerged as a promising tool for innovating how students learn science since late 2022. Despite the buzz around ChatGPT's potential, empirical studies exploring its actual utility in the classroom for learning remain scarce. This study aims to fill this gap by analyzing the lesson plans developed by 29 pre-service elementary teachers from a Korean university and assessing how they integrated ChatGPT into science learning activities. We first examined how the subject domains and teaching and learning methods/strategies were integrated with ChatGPT in the lesson plans. We then evaluated the lesson plans using a modified TPACK-based rubric. We further examined pre-service teachers' perceptions and concerns about integrating ChatGPT into science learning. Results show diverse applications of ChatGPT in different science domains—e.g., Biology (9/29), Chemistry (7/29), and Earth Science (7/29). Fourteen types of teaching and learning methods/strategies were identified in the lesson plans. On average, the pre-service teachers' lesson plans scored high on the modified TPACK-based rubric (M = 3.29; SD = .91; on a 1-4 scale), indicating a reasonable envisage of integrating ChatGPT into science learning, particularly in 'instructional strategies & ChatGPT' (M = 3.48; SD = .99). However, they scored relatively lower on exploiting ChatGPT's functions toward its full potential (M = 3; SD = .93) compared to other aspects. The study also identifies both appropriate and inappropriate use cases of ChatGPT in lesson planning. Pre-service teachers anticipated ChatGPT to afford high-quality questioning, self-directed learning, individualized learning support, and formative assessment. Meanwhile, they also expressed concerns about its accuracy and the risks that students may be overly dependent on ChatGPT. They further suggested solutions to systemizing classroom dynamics between teachers and students. The study underscores the need for more research on the roles of generative AI in actual classroom settings and provides insights for future AI-integrated science learning.

*(Corresponding author: Gyeong-Geon Lee)*
Gyeong-Geon Lee is a Postdoctoral Research Associate at the AI4STEM Education Center, University of Georgia. Address: 110 Carlton Street, Athens, GA 30602 United States (email: crusaderlee@snu.ac.kr; ggleeinga@uga.edu)
Xiaoming Zhai is an Associate Professor at the Mary Frances Early College of Education and the Director of AI4STEM Education Center, University of Georgia. Address: 110 Carlton Street, Athens, GA 30602 United States (email: xiaoming.zhai@uga.edu)

## I. INTRODUCTION

TRADITIONAL teaching and learning paradigms have presumed two intelligent agents in the classroom: a teacher as the primary content deliverer, with students as passive recipients. However, with the advent of information and communication technology around the beginning of the 21st century, numerous scholars anticipated the integration of artificial intelligence (AI) into future classrooms to enhance both teaching and learning [1]. The advancements in machine learning have greatly influenced the evolution of AI. Further, the recent development of generative AI (GenAI) is championed by generative models, such as Generative Pre-trained Transformer (GPT) [2]. GenAI models are usually constructed based on artificial neural networks, trained by large volumes of data to generate new content and data. While GenAI pertains not only to textual data but also to audio-visual ones, the most noticeable growth of GenAI services in real life as of 2023 mostly targets text generation [3]. Large-language models (LLMs) such as ChatGPT have advanced the state of natural language processing, understanding, and generation techniques, bringing innovative educational practices into tangible teaching and learning environments [4][5][6].

The academic community is currently exploring ways to adapt LLMs for educational applications. Notably, ChatGPT has been hailed as a transformative tool for education [4], especially due to its user-friendly nature [6]. Some researchers have proposed the use of ChatGPT in educational settings and outlined potential guidelines and considerations for its deployment [5][7]. Moreover, ChatGPT's potential for revolutionizing science learning has been recognized. For example, Zhai initially reported that ChatGPT could tackle the challenging part of science teaching and learning by automatic assessment development, grading, learning guidance, and recommendation of learning materials [8]. Further, affordable AI services such as ChatGPT could facilitate students' scientific inquiry, supporting them to be Pragmatic Innovators, Foundational Explorers, and Holistic Visionaries when appropriately used with adaptive pedagogical strategies [9].

Likewise, since the release of ChatGPT in November 2022 ignited interest in the educational potential of LLMs [10], the majority of related studies, as of August 2023, have been predominantly visionary and presumable. One important point in those LLM-friendly perspectives of educational scholars is that they are quite optimistic about the universal applicability of LLMs in education. If the visionary scholars are anticipating AI changing how students are provided with personalized guidance, support, and feedback, and teachers and policymakers receiving assistance for instructional decision-making [11], LLMs, like ChatGPT, are said to be close to realizing that ambition [12]. However, Lee et al. noticed that few studies have actually integrated AI into teaching and learning [13], particularly for science subjects, despite the aspirations for AI-driven innovations in education [4][8]. It is quite natural to question how much universality LLMs, specifically ChatGPT, could be used for science education throughout the knowledge domain and teaching and learning methods/strategies. Further, there are limited empirical studies exist that delve into teachers' and students' perceptionsopinions regarding the utility of ChatGPT in science learning. Meanwhile, research has documented conflicted viewpoints on the impact of LLMs on education. For example, there have been critical opinions on the rising interest in LLM in education, different from the abovementioned positive perspectives. These include the potential decrease in students' cognitive engagement [14], possible errors in LLM-generated responses to users' queries [15], ethical issues such as plagiarism [16], and the concern for radically reshaping teaching and learning around technology without careful consideration [17]. While these prudent approaches are worth noting, the early alerts for integrating the new technology should be re-examined based on concrete evidence, given the scarcity of empirical investigations of how ChatGPT could be used for teaching and learning. Therefore, there is a pressing need for empirical research exploring the practical applications of ChatGPT and other LLMs in science education.

To fill the gaps, this study specifically examined how pre-service science teachers adopted ChatGPT in their lesson plans. Analyzing teachers' lesson plans is a pivotal approach to examining teachers' professional competencies and instructional practices [8][18][19][20]. While diverse perspectives exist as the reasons for scrutinizing teachers' lesson plans, recent scholarship has concentrated on how teachers integrate technologies into learning activities [21][22]. Such analyses provide insights into teachers' preparedness and inclination to embed technology within their science curriculum [22], which is particularly meaningful for pre-service teachers, as they are the highly motivated yet less prepared ones in terms of integrating ChatGPT into lesson plans. Specifically, this study examined how upcoming educators destined to instruct future generations incorporate ChatGPT in science lesson plans. Examining their lesson plans, we seek insights into the ongoing academic discourse. Three research questions (RQs) guided this study:

RQ1: What is the applicable range of integrating ChatGPT into science lesson plans by pre-service teachers?

RQ2: How proficient are pre-service teachers in planning and integrating ChatGPT in science learning?

RQ3: What are pre-service teachers' perceived usefulness and concerns regarding integrating ChatGPT into science teaching and learning?

## II. THEORETICAL BACKGROUNDS

### A. Technology Integration in Science Education

While traditional teaching methods dominate the conventional science classroom, efforts to integrate contemporary technologies into science education are underway, leveraging their unique benefits to enhance student learning [23]. From an e-learning perspective, technology acts as a medium, creating distinct impacts on student learning, separating from the content itself [24]. Early concepts like computer-assisted instruction and intelligent tutoring systems, initiated in the 1980s, visualized technology integration in science education. However, the full integration of these technologies was not realized until the 2000s [23]. The growth of ICT hardware and software in the millennium paved the way for more profound technology integration in science education [23], with AI-driven educational initiatives gaining prominence in the 2010s [1][25]. The 2020s witnessed the COVID-19 pandemic, which unintentionally accelerated technology integration in science education as traditional face-to-face methods became impractical, necessitating a technological pivot [26].

Central to the technology integration discourse is identifying barriers and equipping teachers with strategies to incorporate technology into their science curriculum [27][28][29]. Common obstacles from a teacher's perspective include insufficient resources, time constraints, lack of funding, technical issues, and resistance to change, among others [23]. Key to success is a teacher's willingness to use technology and the availability of supportive resources and a conducive environment [30]. This becomes particularly contentious when integrating AI due to concerns about privacy, bias, and surveillance [31][32].

Scholars argue that constructivist learning theories can encourage pre-service teachers' technology integration [28][29]. This sentiment aligns with sociocultural constructivism, emphasizing the role of mediating tools in learning, as suggested by Vygotsky [33]. The focal point of integrating technology in this context is redefining teachers' roles in classrooms where students, empowered by technology, take charge of their learning [27][28], especially with AI tools [1][25][32]. Thus, understanding teachers' perceptions of technology integration becomes a crucial research agenda [27].

### B. Large-language Models and Education

LLMs are a type of AI trained on vast amounts of text data, often referred to as large corpora. Deep neural network-driven LLMs can generate textual responses to users' queries, presenting semantics and syntax in the context of human-AI interactions. Notable LLMs like BERT, XLNet, GPT, LaMDA, and LLaMA have been opened to the public, competing for top performance and dominant market share [34]. Spurred by this open-source approach, researchers are exploring the ideal application of LLMs to foster more equitable education across varying school levels, races, genders, subject domains, and more [7][35]. Specifically, due to their foundational roles as 'language' models, much of the focus has been on their potential applications in language (particularly English) education, such as assisting in essay drafting and refining grammar and style [35].

As highlighted, ChatGPT stands as the state-of-the-art LLM, leading the transformative shifts in anticipated teaching and learning methods worldwide since its debut in November 2022 [4][7][8][35]. Scholars suggest that ChatGPT's versatility in student learning offer capabilities ranging from content creation and profound questioning to idea formulation, essay grading, formative assessment, and more within science education, which are attributed to its exemplary performance in nearly all instructional tasks [36][39]. Notably, while the use of LLMs for automated assessment has been predominant thus far [36][37][38], there is a growing inclination to harness them for tangible educational contexts. In particular, ChatGPT's chatbot-like attributes, which augment the user experience compared to GPT-3.5 and GPT-4, showcase the evolution of LLMs into user-centric learning tools [8]. Also, there are concerns about whether the use of LLMs or ChatGPT could induce damage to authentic learning, as teachers and students might depend too much on it rather than struggling with the tasks themselves [14][15][16][17].

In summary, extensive potential of LLMs has been identified for educational objectives, but there is a lack of empirical studies examining the integration of ChatGPT, or similar LLMs, into science pedagogy. Hence, delving into ChatGPT's potential roles in science classrooms can shed light on the possibilities of LLM in influencing future science education.

*C. Lesson Plan Analysis and Science Teachers' TPACK*

Teachers, whether consciously or subconsciously, design their lesson plans before instruction, implement them, and gather feedback on their teaching. In this context, a lesson plan serves as a vital tool to systematize and model instructional practices [40]. A typical lesson plan comprises learning objectives, materials and resources, learning activities and procedures, assessment, reflection, and more [41]. By developing a comprehensive lesson plan, science teachers establish a framework detailing the teaching and learning processes within the classroom [42]. This structure enables teachers to customize their instruction to suit curricular mandates and the specific needs of their students.

Since lesson plans are crafted prior to the actual teaching, analyzing and revising them based on anticipated classroom situations can considerably bolster a teacher's readiness for instruction [20]. Moreover, an analysis of a lesson plan can unveil a teacher's foundational teaching philosophy and competence. Such a plan encapsulates both the teacher's content knowledge (CK) and pedagogical knowledge (PK). It is worth noting that the technological knowledge (TK) of teachers is increasingly recognized as vital when analyzing lesson plans, especially in the context of technology integration [43]. From this standpoint, researchers in science education have been motivated to create rubrics and assessment tools aimed at evaluating the quality and attributes of pre or in-service science teachers' lesson plans [20][41][44].

The TPACK concept—Technological Pedagogical And Content Knowledge—has emerged as a cornerstone in understanding how teachers view and assess the integration of technology within the classroom [22]. It's important to note that TPACK holistically encompasses CK, PK, PCK (where PK and CK intersect), and TK. The amalgamation of these domains, represented by TPACK, has been proposed in the context of a teacher's lesson plan, especially when they incorporate technology into their science curriculum. Thus, TPACK offers a framework for evaluating a teacher's capability to adeptly, efficiently, and effectively blend technology into science instruction—a capability reflected in their science lesson plans [27]. In line with this, numerous studies have explored pre or in-service teachers' TPACK, probing how it manifests within their lesson plans and real-world teaching practices [42][43][44], as well as developing interventions to enhance it [46][47].

Despite the smattering of research blending AI principles with TPACK [48], there seems to be a lack of studies investigating the TPACK of pre or in-service-service teachers concerning LLMs or ChatGPT.

III. METHOD

*A. Participants*

Twenty-nine pre-service elementary science teachers, comprising 11 males and 18 females, participated in the study. Most of these participants were sophomores at a Korean teachers' university. Under to the compulsory Korean 2015 Revised National Curriculum, they had studied 「Integrated science」, which covers physics, chemistry, biology, earth science, and environmental science, when they were 10[th]-graders. Prior to this study, they underwent a 3-week on-site teacher training experience in schools, which enabled them to experience authentic classrooms. At the time of the study (May-June, 2023), they were enrolled in a "Science Education 1" class taught by the first author. They voluntarily participated in the study and signed the informed consent.

*B. Context*

The instructor, also an author of this paper, conducted four 2-hour sessions with the students, delivering one session each week. The first session introduced and familiarized participants with ChatGPT and its fundamental functions. Most of the pre-service teachers acknowledged that it was the first time they became aware of ChatGPT or any LLMs. The second session discussed the potential benefits of incorporating ChatGPT into science teaching and learning. In this session, the instructor showed examples of making and revising analogies for science concepts using ChatGPT, which either helped elaborate the analogies or deteriorated the quality of analogies in terms of scientific rigor and mapping between the target and the source. In the third session, the instructor detailed various teaching and learning methodologies for science, including the learning cycle,

Predict-Observe-Explain (POE), discovery learning, analogy generation, concept mapping, the epistemological vee, role-playing, etc. The fourth session was dedicated to instructing students on crafting lesson plans, with which all participants had prior experience from their preceding courses. The pre-service teachers proceeded to write their lesson plans during the class hour while the instructor gave feedback and answered the questions from the participants.

*C. Data Collection*

After reflecting on how to weave ChatGPT into the lessons, participants were tasked with drafting a lesson plan suitable for a typical 45-minute elementary science class. They needed to incorporate at least one teaching and learning method they studied in the "Science Education 1" course (or consider alternative methods). The plan was expected to describe and harness ChatGPT's features they experienced for efficient and effective teaching and learning. The prescribed lesson plan format comprised learning objectives/goals, course content, and a lesson outline. It was mandated that in the lesson plan, either the teacher or the students must pose a query to ChatGPT at least once, structuring the lesson around ChatGPT's responses. Additionally, participants were asked to attach a simulated conversation, projecting a potential interaction between the teacher/students and ChatGPT that aligns with their lesson plans. As a resource, they were given access to the national curriculum (mandatory across Korean primary and secondary schools) and elementary school science textbooks.

*D. Rubric*

To understand how the pre-service teachers integrate ChatGPT into their lesson plans, we employed a modified rubric from Harris et al. [22]--TPACK-based Technology Integration Assessment Rubric (TIAR). Harris et al. [22] aimed "to measure the *quality* of" (italic as is) technology integration in teaching. They particularly focused on incorporating TPACK components in the instrument items. TIAR considers four criteria - 'Curriculum Goals & Technologies' stands for curriculum-based technology use, 'Instructional Strategies & Technologies' for using technology in teaching/learning, 'Technology Selection(s)' for compatibility with curriculum goals & instructional strategies, and '"Fit"' for content, pedagogy, and technology togetherness [22], all evaluated at a 4-point scale. Harris et al. [22] tested the TIAR's usability and face validity with dozens of teachers who use technology in teaching, while also requesting them to evaluate 15 example lesson plans. The reliability (Cronbach's $α$) of the four items on the 15 examples were estimated to be .911.

We adapted this rubric to align with our study focus. For instance, since our study was contextualized for ChatGPT as a specific technology to be integrated into lesson plans, we replaced "technology" with "ChatGPT" to avoid confusion for the participants. Also, we revised the third aspect as 'ChatGPT function selection(s).' By the 'function' of ChatGPT, we denoted the affordance that ChatGPT could provide a student with responses to the student's specific query. Therefore, 'function selection(s)' means how well the pre-service teacher designed the use of ChatGPT to induce its potential to support student learning. Consequently, the four criteria now correspond to 'curriculum goals & ChatGPT,' 'instructional strategies & ChatGPT,' 'ChatGPT function selection(s),' and 'fit' (see Table 1). In essence, the lesson plans of the pre-service science teachers were evaluated based on these four TIAR-derived items. The reliability (Cronbach's $α$) of the four items used in this study was estimated to be .69, based on the results for 29 lesson plan evaluation, which falls into the range of 'acceptable' values [49].

--- Insert Table 1 here ---

*E. Survey*

To further understand how the participating pre-service teachers determined the content and teaching methods, as well as their perceived usefulness of ChatGPT and the anticipated outcomes, we asked them to complete a survey, encompassing three open-ended questions: (1) Rationale behind the selection of specific content and teaching method/strategy. (2) Reasons for deeming ChatGPT's integration beneficial for the chosen content and method/strategy. (3) Anticipated outcomes of melding ChatGPT with the selected content and method/strategy, viewed through the lens of the overarching goals of science education.

*F. Analysis*

Two researchers independently coded 10 and 9 science lesson plans respectively based on TIAR and the survey responses. Preliminary discussions were conducted centering on establishing a consensual understanding of the analytical framework between the two researchers. While the majority of analyses aligned, occasional discrepancies happened and were settled through iterative discussions. For the survey data, we adopted the constant comparative method [50] as our primary tool for category derivation. We initiated our analysis with a subset of the data to determine initial categorization. Subsequently, the remaining data portions were reviewed to verify the categorization's overarching applicability. Category modifications were made as and when emergent categories were identified. This iterative process continued until the entirety of participants' survey results were aptly analyzed using the formulated categories. The chat data between the participants and ChatGPT, which were required to be attached to each lesson plan to exemplify the simulated use of ChatGPT in the teaching and learning process, were utilized for triangulation purposes.

IV. RESULTS

In the results section, we first provide a descriptive overview of how ChatGPT was integrated into pre-service teachers' science lesson plans. We then present the TIAR-based scores of pre-service teachers' lesson plans that incorporated ChatGPT, complemented by example lesson plans. Lastly, we qualitatively report on pre-service teachers' ~~perceptions~~thoughts regarding the integration of ChatGPT into science teaching and learning.

*A. How ChatGPT was Integrated in Science Lesson Plans by Pre-service Teachers*

We analyzed a total of 29 lesson plans to understand how pre-service teachers incorporated ChatGPT in learning or instructional activities. In this subsection, we descriptively present how various knowledge domains and teaching and learning methods/strategies were used in their lesson plans.

We found that the pre-service teachers targeted five knowledge domains with varying percentages: biology (9/29), chemistry and earth science (7/29 each), and physics and environmental science (4/29 each). It should be noted that participants could select multiple knowledge domains if the content spanned cross-cutting concepts.

In terms of teaching and learning methods/strategies, pre-service teachers predominantly employed cooperative learning (7/29), POE or the learning cycle (6/29 each), analogy generation (4/29), experimentation (3/29), and concept mapping (3/29), among others.

A summary of the target knowledge domains is presented in Figure 1-(a), and the teaching and learning methods/strategies can be found in Figure 1-(b). Summaries of sample lesson plans that integrate ChatGPT are shown in Table 2. To show how teachers integrate ChatGPT in the lesson plans, we present one exemplary case from *participant_11* in Figure 2 (see Appendix A for the English version of Figure 2).

The *participant_11*'s lesson plan was structured around teaching strategies for $5^{th}$-grade students, focusing on the theme "Life and Environment," particularly "Disruption of Ecological Balance." It was structured into several sections: objectives, materials, classroom strategies, and evaluation methods. The objectives are clearly listed, aiming to impart knowledge about ecological balance, its importance, and the consequences of its disruption. The materials section outlines the resources needed for the module, including videos, role-play scripts, and interactive tools like ChatGPT. The classroom strategy section is subdivided into different phases, detailing the activities, time allocation, and teacher-student interactions for each phase. This includes introductory activities, main lesson development with interactive elements like role play, and concluding activities with a focus on reinforcement and evaluation. The evaluation section provides criteria for assessing student understanding and engagement through the activities.

In *Participant_11*'s lesson plan, ChatGPT was integrated for fourfold purposes: (1) ChatGPT was proposed to let "one can go beyond the content presented in textbooks and learn about additional reasons for the disruption of ecological balance," while the content matter of ecological balance is "often briefly covered and overlooked in traditional textbook learning" ($1^{st}$ aspect of TIAR). (2) ChatGPT was envisioned to support students' role-playing activity, as like "in various preparation stages such as setting up role-play characters, establishing the overall direction of the content, and adjusting the timing of the role-play, one can receive assistance from ChatGPT" ($2^{nd}$ aspect of TIAR). (3) ChatGPT was anticipated to provide each group or individual "adaptive instruction," via further "explorative questions or additional information" ($3^{rd}$ aspect of TIAR). (4) ChatGPT could deliver feedback and reflection on "whether the role-play has been well-structured according to your intentions, what aspects were executed well, and what aspects are lacking and need improvement" ($4^{th}$ aspect of TIAR). To sum up, *participant_11* intended to use ChatGPT to aid students in researching and understanding the disruption of ecological balance beyond textbook content, assist in scriptwriting for role plays, enhance individual and group activities, and promote feedback and reflection. *Participant_11*'s case shows that pre-service teachers earnestly deliberated how they could integrate ChatGPT into their lesson plans, exploiting its potentials for supporting science learning.

--- Insert Figure 1 here ---
--- Insert Table 2 here ---
--- Insert Figure 2 here ---

*B. Strengths and Weaknesses of ChatGPT-integrated Lesson Plan by Pre-service Teachers*

We elucidated the characteristics of the pre-service teachers' lesson plans in the subsequent section, referencing scores based on the TIAR. For example, *participant_11*'s lesson plan was evaluated to have 4 points for all four aspects of TIAR. Table 3 provides an evaluation of the science lesson plans devised by all the pre-service teachers that integrated ChatGPT. The average TIAR score for these plans was M = 3.29 (SD = .91), exceeding the mid-point of 2.5 on a 4-point scale. This result suggests that pre-service teachers possess an averagely high level of TPACK competency for integrating ChatGPT into science education. The lesson plans were scored an average of M = 3.44 (SD = .87) for 'curriculum goals & ChatGPT' (item 1), M = 3.48 (SD = .99) for 'instructional strategies & ChatGPT' (item 2), M = 3 (SD = .93) for 'ChatGPT function selection(s)' (item 3), and M = 3.24 (SD = .83) for 'fit' (item 4). A paired samples $t$-test was conducted to compare differences in the scores of items 1-4. There was a significant difference only in the scores for item 2 and item 3; $t(28) = 2.3173$, $p = .0280$ ($< .05$). This indicates that pre-service teachers are more competent in incorporating ChatGPT into their instructional strategies than in selecting the most appropriate ChatGPT functions for their lesson plans.

--- Insert Table 3 here ---

Given that 'fit' (item 4) pertains to the overall coherence of the first three items, it is insightful to examine items 1-3, specifically to explore why 'ChatGPT function selection(s)' (item 3) registered a lower score than items 1-2.

In terms of 'curriculum goals & ChatGPT' (item 1), pre-service teachers aimed to harness ChatGPT to aid students in acquiring cognitive and "conceptual knowledge," a pivotal curriculum objective in (Korean) science education. It is noteworthy that a diverse range of science concepts, such as the 'solar system and stars,' 'acids and bases,' 'life and environment,' and 'lives of animals,' were integrated into the lesson plans (Table 1). Additionally, cognitive educational goals like "scientific inquiry," "scientific thinking," and "creative thinking," as well as affective goals such as "motivation," "interest," and "attitude," were aligned with the use of ChatGPT. All the abovementioned curriculum objectives/goals quoted from pre-service teachers' lesson plans were also existent in Korean 2015 Revised National Curriculum document.

Regarding 'instructional strategies & ChatGPT' (item 2), various teaching and learning methods/strategies were outlined by the pre-service teachers (Figure 1-b). Cooperative learning emerged as the most common instructional strategy (7/29). For instance, it was proposed that ChatGPT could support expert group learning within the Jigsaw cooperative learning method. In the context of POE (6/29), ChatGPT was seen as beneficial for aiding students' prediction or explanation processes. Similarly, for the learning cycle (6/29), ChatGPT was perceived as a tool to facilitate students' exploration of inquired phenomena. Analogy (4/29), being fundamentally a language-based task, was also frequently recommended, as ChatGPT could be particularly useful in this regard.

In the case of 'ChatGPT function selection(s)' (item 3), we deduced both the appropriate and inappropriate applications of ChatGPT in science lesson plans from the 29 samples. Some pre-service teachers showcased appropriate usage of ChatGPT (Table 1). For instance, *Participant_7* viewed ChatGPT as a supplementary information source alongside teachers and textbooks. *Participant_11* believed ChatGPT could assist in organizing role-playing activities, specifically in defining roles, direction, and time management. ChatGPT's ability to distill lengthy student-generated scenarios into concise summaries, appropriate for a 45-minute class duration, was particularly noted, underscoring the inherent strengths of ChatGPT as a language model. *Participant_20* highlighted ChatGPT's capability for supporting extended, consecutive questioning. He also offered a strategy to temper students' over-reliance on ChatGPT by structuring sessions within a cooperative learning framework.

Nevertheless, there were instances of inappropriate uses of ChatGPT (Table 1). For example, *Participant_3*'s lesson plan employs the POE method to teach about the solar system and stars. The participant suggested that the teacher guide students to find constellation's locations, which have a visual nature, inapt to use ChatGPT as it does not afford much visual information (as of May 2023). *Participant_20* attempted to procure additional resources, like YouTube video links on animal lives, based on ChatGPT's recommendations. It was misguided, as the YouTube video links provided by ChatGPT turned out to be non-existent. Also, *participant_20* submitted the ChatGPT-generated video summaries to support her rationale for the lesson planning, which falls short because the video does not exist (it is notable that the participant did not even check whether they exist). *Participant_8* suggested that student groups use ChatGPT to design experiments on acids and bases. This use needs particular caution as it could potentially undermine students' practical inquiry skills over time.

In conclusion, the varied scores for 'ChatGPT function selection(s)' can be attributed to the mix of both suitable and unsuitable applications of ChatGPT within the lesson plans.

*C. Pre-service Teachers' Perceived Usefulness and Concerns of Integrating ChatGPT in Lesson Planning*

Qualitative analysis of teachers' responses to the questionnaire uncovers that pre-service teachers expressed their anticipations regarding the contributions that ChatGPT could make to science teaching and learning. The following points were closely interrelated in the pre-service teachers' lesson plans and responses to the survey but logically differentiate for the ease of report (see Table 4 for the summary).

---Insert Table 4 here ---

First, a majority recognized that ChatGPT could revolutionize the manner in which students construct knowledge through questioning (16/29 participants). Given its capability to offer an "instant answer" to any user query (*participant_6*), its affordance was linked to the self-paced nature of learning. As such, it could enable students' "active participation" (*participant_12*) and assist them in "managing and developing their own learning process" (*participant_7*).

Pre-service teachers especially valued ChatGPT's context-awareness in the dialogue with the user, which is a multi-turn conversation. For example, "when I want to obtain more in-depth information on a given topic, multiple searches are often necessary. By using ChatGPT, I can get the answers I'm looking for through continuous questioning" (*participant_20*). They believed that using ChatGPT might be superior to traditional internet searches for information retrieval. This is because ChatGPT allows users to pose consecutive questions and summarizes the answers sticking to the central topic, in contrast to the vast amount of information presented by standard internet searches (3/29 participants). Further, since this capability was characterized as consecutive 'questioning' being not satisfied by the given answers, which is impossible when only textbook is used, pre-service teachers considered ChatGPT as a candidate to bolster students' critical thinking (5/29 participants).

Significantly, they believed that ChatGPT could offer individualized learning support for each student, a task that a single teacher cannot fulfill independently (9/29 participants). Hence, ChatGPT was perceived as a tool that could "enable individualized and self-directed learning" (*participant_19*). The self-directed learning facilitated by ChatGPT was also expected to boost students' "motivation" and "interest."

Another noteworthy potential application of ChatGPT that pre-service teachers highlighted was for formative evaluation (8/29 participants). For instance, ChatGPT could offer students "evaluation" or "feedback" in activities like role playing (*participant_11*), information searching (*participant_16*), and prediction or explanation (*participant_17*). Interestingly, a couple of pre-service teachers imagined that contrasting students' answers with ChatGPT's responses to specific questions (e.g., 'how is fried ice cream possible?' - *participant_5*) could enhance the depth of student answers.

Additionally, some pre-service teachers (4/29) contended that the integration of ChatGPT is inherently beneficial as it fosters students' "digital competency [or literacy]" in preparation for a society increasingly gravitating towards AI-centric industries (3/29).

Yet, pre-service teachers also expressed concerns about integrating ChatGPT into science learning, especially when limited evidence about accuracy, reliability, and security has been provided. Primarily, they were concerned about the precision of ChatGPT's responses and its reliability (11/29). Given that ChatGPT's "training data extends only up to September 2021 and doesn't encompass the latest scientific advancements, it can occasionally deliver inaccurate or incorrect information" (*participant_8*). Another teacher underscored potential biases, noting that ChatGPT "might sometimes provide misleading or incorrect information" (*participant_11*). There were particular concerns that students might develop "misconceptions" due to ChatGPT (*participant_13*), and the ramifications would be severe if ChatGPT delivered incorrect information about lab safety (*participant_24*). In consequence, the pre-service teachers were also worried that students might be overly dependent on ChatGPT, given its feasibility and high accessibility (13/29).

Notably, almost every pre-service teacher (11/29) who raised the issue of over-dependence on ChatGPT (13/29) also proposed remedies that could be described in the two interrelated aspects. The first aspect to counteract it was to diversify information sources instead of solely relying on ChatGPT. Table 5 shows the other learning sources pre-service teachers mentioned, and the number of teachers supported each. The most important learning source was the teacher (3/11). For example, "teacher should instruct students to maintain critical thinking and compare ChatGPT's answer to other information." (*participant_8*). Also, "teachers listen to students' presentation and correct misconceptions while providing students with feedback" (*participant_14*). In addition, students themselves (2/11) or their peers (1/11) were found to be the additional information source. The unspecified "diverse information" (1/11) might include textbooks and traditional internet searching. Interestingly, a couple of pre-service teachers (2/11) said that ChatGPT could provide further and diverse information, adding to what it had provided before (e.g., "if [teachers] use ChatGPT that teaches detailed information on diverse viewpoints …" - *participant_15*; "Search additional cases using ChatGPT … could lead to constructing correct concept and resolving the over-dependence problem" - *participant_18*). This shows that some pre-service teachers are thinking of ChatGPT as an affordable for diverse, balanced, and self-correcting learning sources.

--- Insert Table 5 here ---

For the second aspect, pre-service teachers have developed systematic methods to structure learning processes to reduce students' over-dependence on ChatGPT, emphasizing classroom interactions (11/29). Table 6 shows students' concern about using ChatGPT in science classrooms and their solutions to the problem -- before, during, and after using ChatGPT. For example, *participant_14* recommended educating students about the "characteristics of ChatGPT," "digital literacy," and "critical thinking" before using it in the classroom. *Participant_5* suggested that "[structured] group discussions can effectively manage" the excitement generated while using ChatGPT in the classroom, thereby maintaining a focused learning environment. Students also highlighted the importance of reflection after using ChatGPT. For instance, *Participant_10* suggested that although "one can ask ChatGPT about their predictions" in the prediction stage of the "POE model", "it is essential to conduct experiments to verify whether your prediction was correct or not."

--- Insert Table 6 here ---

## V. DISCUSSION

In this study, we provided professional learning about ChatGPT to 29 pre-service teachers and examined how they incorporate ChatGPT in the learning and instruction activities and their perspectives on ChatGPT in science teaching and learning. We found that pre-service science teachers envisioned the extensive applicability of ChatGPT in teaching and learning, were moderately competent in integrating ChatGPT in the lesson plan, and showed balanced perspectives on the merit of integrating ChatGPT for science education The findings have several conceptual and practical implications.

### A. Pre-service teachers envisioned the extensive applicability of ChatGPT in science teaching and learning (RQ1)

First, upon analyzing the lesson plans crafted by pre-service teachers, it became evident that they see ChatGPT as having a robust potential for a wide array of applications in diverse science teaching and learning scenarios (RQ1). As anticipated, these teachers adeptly incorporated ChatGPT into teaching a spectrum of science content spanning domains like biology, chemistry, earth science, physics, and environmental science. What is remarkable, however, is the fact that they amalgamated ChatGPT with 14 teaching and learning methods/strategies, which include cooperative learning, POE, learning cycle (6/29 each), analogy generation (4/29), concept map (3/29), and experimentation (3/29).

These visionary lesson plans crafted by pre-service teachers suggest that they ~~both current and future in-service teachers~~ are likely to utilize ChatGPT in future science classrooms, merging various combinations of content knowledge with diverse teaching and learning methods. Two contrasting discussion points arise here:

(1) Further explorations into how ChatGPT can be applied in science teaching and learning are imperative. While participants in this study primarily envisioned using ChatGPT in a face-to-face setting, opportunities for remote or non-face-to-face learning that incorporates ChatGPT shouldn't be overlooked. Moreover, the development of innovative science teaching strategies that leverage ChatGPT is essential.

(2) At the same time, we must critically assess whether there truly are no obstacles when it comes to integrating ChatGPT into various science teaching formats. Despite the majority of pre-service teachers in this study anticipating limitless integration of ChatGPT into their lessons, potential challenges should be acknowledged. For instance, even if barriers related to "behavior, investments, and commitment of individual teachers" are minimal, issues such as "technical problems, … poor administrative support, [and] poor training" might remain [23]. Therefore, identifying the potential barriers to integrating ChatGPT in science classrooms merits consideration for future research.

### B. Pre-service teachers were moderately competent in integrating ChatGPT as technology for teaching and learning (RQ2)

Second, discrepancies were evident in the pre-service teachers' TPACK components, as demonstrated in their lesson plans (RQ2). They displayed proficiency in 'curriculum goals & ChatGPT' (3.45/4) and 'instructional strategies & ChatGPT' (3.48/4) yet struggled with 'ChatGPT function selection(s)' (3/4) and 'fit' (3.24/4). The average score was 3.29/4, which is higher than the mid-point of 2.5.

Given that these pre-service teachers are future educators in classrooms, the traits observed in their lesson plans will likely be indicative of future approaches to ChatGPT-integrated teaching. Consequently, this study's findings underscore the need for enhanced professional development for science teachers in AI integration within science education, which aligned with prior research [51]. A qualitative investigation of how teachers choose and arrange content, instructional strategies, and particularly ChatGPT functions is crucial. By identifying both strengths and weaknesses in the actual design and implementation of ChatGPT-integrated lessons by both pre- and in-service teachers, we could devise targeted support measures. As this study reveals, there's a limited understanding among pre-service teachers regarding ChatGPT functionalities, necessitating professional development to prevent potential challenges in the classroom. This observation points to the need to create professional development programs focusing on incorporating ChatGPT and similar LLMs in science education.

### C. Pre-service teachers showed balanced perspectives on the merit of integrating ChatGPT for teaching and learning (RQ3)

Third, the insights from pre-service teachers on incorporating ChatGPT into science lessons offer a preview into the evolution of future science instruction, resonating with existing AIEd literature [1][25] (RQ3).

With its prompt feedback and contextual sensitivity, ChatGPT is poised to revolutionize student questioning, paving the way for more self-directed and personalized learning—options previously constrained by teacher-student ratios [27][28]. Some even advocate for ChatGPT's potential in formative assessment within science classrooms [36], hinting at the advent of an automated real-time feedback system.

Notably, these pre-service educators recognized and addressed potential challenges of ChatGPT integration. They underscored the importance of diversifying information sources like teachers and peers to counter possible misinformation from ChatGPT. This perspective aligns with the growing emphasis on critical thinking and digital literacy in the LLM era [5][7]. Most significantly, to counteract possible over-dependence on ChatGPT, they proposed structured classroom interactions, leveraging established educational strategies like group discussions, POE, and student-led presentations with teacher feedback. In essence, fostering active teacher-student and peer interactions during structured learning can temper undue student-AI dependencies. Further concretization of the AI-integrated science instruction strategies, as well as a thorough investigation of classroom dynamics before and after introducing LLMs, is essential.

### D. Limitations and Future Directions

The primary limitation of this study stems from its reliance on pre-service teachers' lesson plans as data sources. Given the potential discrepancies between planned and executed lessons, it's crucial to further empirically examine the utilization and perceived usability of ChatGPT and other LLMs in real-world science classrooms.

While this study found the promise of ChatGPT, it is critical to realize the cost of ChatGPT (subscription fee), which may not be affordable for many underresourced teachers and students. Future studies might explore how much open-source LLMs, such as Llama, Bloom, and Falcon, could also contribute to equitable and transparent practices of AI-empowered teaching and learning.

As of May-June 2023, when the data was collected for this study, there were few prompt engineering techniques developed and spread to broader users. As of January 2024, it is deemed possible to develop LLM-based tools tailored to support teaching specific scientific content with designated instructional methods. Therefore, the effort from the technical side to assist teachers who are willing to integrate ChatGPT and other LLMs is also called.

Additionally, exploring strategies to prevent teachers' excessive reliance on ChatGPT or other GenAI models (e.g., Google's Bard) during their instructional design and implementation is needed. As exemplified in this study, GenAI models could present incorrect or even non-existent information/materials as they sometimes hallucinate their response, and teachers might accept it without checking. Critical thinking or digital literacy on what ChatGPT provides will be important expertise of future science teachers.

Finally, research should delve into the impacts of ChatGPT-integrated science instruction from the student perspective. It's crucial to ascertain if lessons incorporating ChatGPT genuinely amplify students' cognitive gains in science, enhance their scientific inquiry abilities, and positively influence their attitudes toward science. In this regard, although this study provides empirical evidence, carefully designed quasi-experimental studies to compare experimental and control groups should follow to estimate the impact of integrating ChatGPT on science teaching and learning.

*E. Implications*

The broad implication of this study pertains to the use of GenAI models or LLMs in education. This study appears to be among the initial efforts exploring the integration of ChatGPT into science lesson plans, which can be applied in actual science classrooms. Given the lack of empirical research on the potential introduction of ChatGPT or other LLMs in science education, this study can offer insights for science education researchers and practitioners. One of those insights is the importance of equipping teachers and students alike with digital or AI literacy so that they discern and use what is suitable for learning among the content that GenAI generates.

This study sketches the configuration of future science classrooms, where AI will bring substantial changes. As evidenced by the pre-service teachers' lesson plans, GenAI -- more specifically, ChatGPT, could enhance personalized learning, help teachers manage classrooms with a large number of students, and enable students to proceed with interactive learning activities, which are much more fascinating compared to the traditional classrooms. Also, it seems that ChatGPT can be integrated with almost every combination of knowledge domains and instructional methods/strategies, which shows its scalability. Future studies should design and develop instructional programs with integrated ChatGPT for science education, in line with this study, which could lead the way to transforming the classroom interactions between teachers and students, mediated by AI, at least partially.

Most significantly, this study reported that pre-service teachers were not only aware of the potential pitfalls of integrating ChatGPT into science classrooms but also devised their own ways to remedy students' over-dependence on it. For example, they suggested double-checking information given by ChatGPT with various other sources and structuring classroom interactions to systematically manage students' use of ChatGPT. Notably, the solutions to the ChatGPT's pitfalls were not required by the instructor. However, the pre-service teachers voluntarily provided how they contemplated the uses of ChatGPT. This shows a glimpse of future teachers' responsive TPACK to the changing educational circumstances by AI, maintaining their directing role in the classrooms. Teachers prepared with TPACK are essential even in the era of AI.

--- Insert Appendix A here ---

Table 1. The Technology Integration Assessment Rubric (TIAR)-based rubric to assess ChatGPT-integrated science lesson plans (revised from Harris et al. [22])

| | Item | 4 | 3 | 2 | 1 |
|---|---|---|---|---|---|
| 1 | Curriculum goals & ChatGPT | ChatGPT used in the instructional plan are **strongly aligned** with one or more curriculum goals. | ChatGPT used in the instructional plan are **aligned** with one or more curriculum goals. | ChatGPT used in the instructional plan are **partially aligned** with one or more curriculum goals. | ChatGPT used in the instructional plan are **not aligned** with one or more curriculum goals. |
| 2 | Instructional methods/strategies & ChatGPT | ChatGPT use **optimally supports** instructional methods/strategies | ChatGPT use **supports** instructional methods/strategies | ChatGPT use **minimally supports** instructional methods/strategies | ChatGPT use **does not support** instructional methods/strategies |
| 3 | ChatGPT function selection(s) | ChatGPT function selection(s) are **exemplary**, given curriculum goal(s) and instructional methods/strategies. | ChatGPT function selection(s) are **appropriate, but not exemplary**, given curriculum goal(s) and instructional methods/strategies. | ChatGPT function selection(s) are **marginally appropriate**, given curriculum goal(s) and instructional methods/strategies. | ChatGPT function selection(s) are **inappropriate**, given curriculum goal(s) and instructional methods/strategies. |
| 4 | 'Fit' | Content, instructional methods/strategies and ChatGPT function **fit together strongly** within the instructional plan. | Content, instructional methods/strategies and ChatGPT function **fit together** within the instructional plan. | Content, instructional methods/strategies and ChatGPT function **fit together somewhat** within the instructional plan. | Content, instructional methods/strategies and ChatGPT function **do not fit together** within the instructional plan. |

Table 2. Summaries of Example Pre-service Teachers' Lesson Plans that Integrate ChatGPT

| Participant No. | Gender | Content | Teaching method/strategy | Use of ChatGPT |
|---|---|---|---|---|
| 3 | M | Solar system and stars | POE | - Students ask ChatGPT how the constellations are positioned in the night sky. |
| 7 | F | Solar system and stars | Jigsaw cooperative learning | - Students use ChatGPT as a source of information, which is complementarily used with others (teacher, textbook, etc), in expert group activity of Jigsaw cooperative learning. |
| 8 | F | Acid and base | Questioning, discussion, cooperative learning, experimentation, discovery learning | - Students ask ChatGPT what 'first egg' means.<br>- Students design group experiment using ChatGPT. |
| 11 | F | Life and environment | Role playing | - Students get help from ChatGPT organizing role playing activity (role composition, directing, timekeeping, scenario summarization)<br>- Students are provided with further content knowledge that are not presented in the textbook, by ChatGPT. |
| 20 | M | Lifes of animals | Learning cycle + Cooperative learning | - Students repeatedly ask ChatGPT in a chain-of-thoughts sense.<br>- The structured cooperative learning model would mitigate students' excessive interest in using ChatGPT.<br>- Students get information sources such as YouTube video links from ChatGPT. |

Table 3. The scores of pre-service teachers' ChatGPT-integrated lesson plans according to the revised TIAR rubric

| | Item | Mean | SD |
|---|---|---|---|
| 1 | Curriculum goals & ChatGPT | 3.44 | .87 |
| 2 | Instructional methods/strategies & ChatGPT | 3.48 | .99 |
| 3 | ChatGPT function selection(s) | 3 | .93 |
| 4 | 'Fit' | 3.24 | .83 |
| | Overall | 3.29 | .91 |

Table 4. Main themes from the qualitative analysis and the number of participants agreeing to each ($N = 29$)

| Theme - Subtheme | Number of participants agree |
|---|---|
| ChatGPT could revolutionize the manner in which students construct knowledge through questioning | 16 |
| - ChatGPT is superior to traditional internet search, as it allows consecutive questions and summarizes answers | 3 |
| - Questioning to ChatGPT would foster students' critical thinking | 5 |
| ChatGPT could offer individual learning support for each student | 9 |
| ChatGPT can be applied for formative assessment | 8 |
| ChatGPT is beneficial as it fosters students' digital competency/literacy | 4 |
| Concerns about ChatGPT's accuracy, reliability, and security | 11 |
| Concerns about students' over-dependence to ChatGPT | 13 |
| - Suggestion of systematic methods to reduce students' over-dependence | 11 |

Table 5. Pre-service science teachers' suggestion of other information sources to reduce students' over-dependence on ChatGPT (*n* = 11)

| Information source | Number of participants suggested |
|---|---|
| Self | 2 |
| Peer | 1 |
| Teacher | 3 |
| Experiment | 1 |
| ChatGPT | 2 |
| Unspecified | 1 |

Table 6. Pre-service science teachers' systematic learning process to reduce students' over-dependence on ChatGPT ($n = 11$)

|  | Systematic method | Number of participants suggested |
|---|---|---|
| Before using ChatGPT | Teach students the characteristics and possible pitfalls of ChatGPT. | 6 |
| During using ChatGPT | Structure the learning process so that students use ChatGPT limitedly, and have enough time to interact each other and with teacher. | 6 |
| After using ChatGPT | Verify the information ChatGPT provided with students. | 8 |

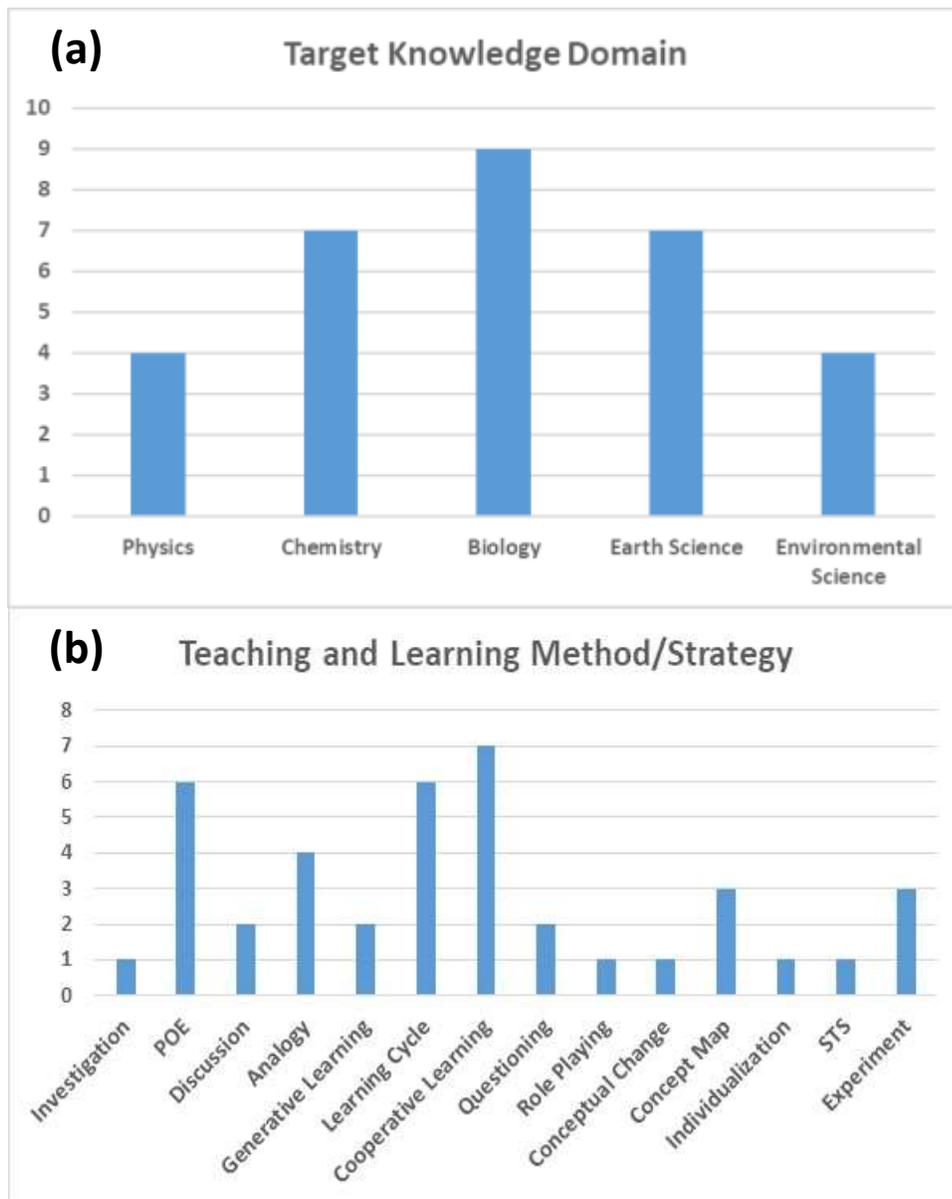

Fig. 1. (a) Target knowledge domain in the lesson plans (b) Teaching and learning method/strategy used in the lesson plans

Fig. 2. Example (clip) of a Pre-service Teacher's Lesson Plan that Integrated ChatGPT (*participant_11*) (See Appendix A for the English version)

Appendix A. Example (clip) of a Pre-service Teacher's Lesson Plan that Integrated ChatGPT (*participant_11*)

## Teaching and Learning Plan

| Unit | 5th Grade Unit 2, 02. Life and Environment | Teaching and Learning Strategy | Role Play |
|---|---|---|---|
| **Main Topic** | Disruption of Ecological Balance | **Duration** | 40 minutes |
| **Learning Objectives** | • Utilize ChatGPT to explore knowledge related to the disruption of ecological balance.<br>• If necessary, use ChatGPT to write a script for a role play depicting scenarios of ecological balance disruption.<br>• Realize the importance of various organisms living within the ecosystem. | | |
| **Materials** | • Smart devices, Class flow PPT, Activity sheets, Writing tools (and various other tools needed for the role play) | | |

| Stage (Duration) | Learning Elements | Teaching - Learning Process | Duration (Minutes) | Notes |
|---|---|---|---|---|
| **Introduction (7 minutes)** | Class Preparation, Start | ○ Preparing for the class<br>○ Starting the class | | · Utilize a predetermined class slogan to foster concentration.<br>· Prepare smart devices fully charged. |
| | Recalling Prior Learning | ○ Recalling prior learning<br>- Students present quizzes based on the content learned in the previous class.<br>· Before starting today's class, let's solve some quizzes based on what we learned last time. Is there anyone who wants to present a quiz?<br>- The student who presents the quiz proceeds with the explanation.<br>· Last time, we learned about the ecological pyramid and ecological balance, right? Someone posed a related question, could you explain it once more, OO? | 2 | · Use quizzes created by students to recall the content of prior learning (to allow the lesson's content to emerge from the students' own words).<br>· Allow students to explain the quiz themselves, providing them an opportunity to think. |
| | Motivating Learning | ○ Motivating learning<br>- Watch a video on a case of ecological balance disruption to encourage students to think about the issue and to induce learning motivation and problems.<br>· Check the content of the video & ask thought-provoking questions.<br>ex) What caused the ecological destruction in Australia shown in the video? How did they try to solve it? What was the result? What should have been done? What should be done in the future? | 4 | · Prepare the video in advance - to be included in the PowerPoint https://www.youtube.com/watch?v=GcqOUxn4uQ8<br>· Create an atmosphere where students can freely express their thoughts.<br>· Summarize the content of the video based on the students' presentations and lead to the identification of learning problems. |
| | Presenting Learning Problems and Learning Activities | ○ Identifying the learning problem<br><br>Let's use ChatGPT to write a script for a role play that illustrates the disruption of ecological balance. | 1 | |

| | | ○ Guiding the learning activities<br><br><Activity 1> Explore knowledge related to the disruption of ecological balance using ChatGPT<br><Activity 2> Write a script for a role play on ecological balance disruption using ChatGPT | | |
|---|---|---|---|---|
| **Development (27 minutes)** | Activity 1 | <Activity 1> Explore knowledge related to ecological balance disruption using ChatGPT (individual activity)<br><br>○ Everyone should summarize the causes of ecological balance disruption they already know on an activity sheet.<br>○ Use ChatGPT to further explore various causes of ecological balance disruption and summarize them on the activity sheet.<br>(This will help in obtaining a variety of answers and learning about causes beyond the textbook to broaden perspectives).<br><br>○ Think of solutions to these causes and summarize them on the activity sheet.<br>○ Use ChatGPT to learn about solutions for each cause and summarize them on the activity sheet.<br><br>○ Decide on a specific cause of ecological balance disruption to focus on for the role play and delve deeper into it. | 8 | · Ensure students are already taught how to use ChatGPT and helpful prompts for utilization.<br>· Utilize the activity sheet to separate existing knowledge from the information provided by ChatGPT.<br>· Guide students to first think of causes and solutions themselves before using ChatGPT to avoid over-reliance.<br>· Continuously provide feedback on students' activities and offer additional explanations as needed (circulating guidance is crucial. Repeatedly check if the information provided by ChatGPT is accurate and aligns with scientific understanding).<br>· Continually engage in questioning that stimulates thinking. |
| | Activity 2 | <Activity 2> Write a script for a role play on ecological balance disruption using ChatGPT (group activity)<br><br>○ Share individually researched causes and solutions, and collectively decide on the content for the role play within the group.<br>- Everyone in the group takes turns presenting.<br>- Choose the main cause, consequences, and solutions to be addressed in the role play.<br>○ Write a brief script illustrating a situation of ecological balance disruption for the role play.<br>- The role play should last no more than 3 minutes (quality over quantity, it should not be too long).<br>- Plan a detailed ecological balance disruption scenario (cause, result, solution).<br>- Analyze necessary roles and select actors for the role play.<br>- When needed, use ChatGPT to get ideas or to supplement, refine, or revise the | 19 | · Transition into group formation to facilitate the activity.<br>· Guide students to collaborate and solve the assigned tasks.<br>· Instruct students on important considerations for role-playing.<br>· Circulate to ensure all groups are progressing smoothly with their activities.<br>· Ensure the role play is not merely considered entertainment but a learning activity.<br>· Encourage maximum discussion among group members to generate creative ideas, minimizing dependency on ChatGPT.<br>· If the role play script is extensive, guide the students to request ChatGPT to summarize long texts, enabling them to complete a script of suitable length for the |

| | | content (for example, using a script example made by ChatGPT to guide the activity).<br>- (For groups that finish early) Assign the task of creating props needed for the role play and practicing the role play. | | given time. |
|---|---|---|---|---|
| **Summary (6 minutes)** | Summarizing Learning Content | ○ Summarize learning content<br>- Discuss and reflect on what was learned and felt during the session.<br>- Share experiences of using ChatGPT, including the positives, challenges, and what was found useful. | 2 | · Guide students to share detailed feedback (Ensure to follow up with further questions if responses are too simple like "it was good" or "it was hard"). |
| | Evaluating | ○ Evaluate<br>- Get feedback on the role play script from ChatGPT.<br>- Provide the script to ChatGPT and receive feedback on whether the role play is well composed according to the intention, what aspects were well done, what needs improvement, etc.<br>- Explain and reflect on the process of how the script was composed, what concepts were utilized, etc. | 3 | · Use ChatGPT to assess the extent of students' learning achievements.<br>· Be cautious of over-reliance on ChatGPT for evaluations; teachers should also participate in the assessment. |
| | Previewing the Next Session | ○ Preview of the next session<br>· Discuss the attitudes to be adopted going forward after presenting the role plays based on the scripts created by each group.<br>○ Assign homework<br>· Refine the role play script, practice presenting the role play. | 1 | |

10